\documentclass[amsmath,amssymb,11pt]{article}
\def\d{\mbox{\rm d}}

\def\cosec{\mbox{\rm cosec}}

\def\half{\mbox{$\frac{1}{2}$}}
\def\dddot#1{\mathinner{\buildrel\vbox{\kern5pt\hbox{...}}\over{#1}}}
\def\ddddot#1{\mathinner{\buildrel\vbox{\kern5pt\hbox{....}}\over{#1}}}

\title{Lagrangians Galore}
\author{M.C. Nucci and P.G.L. Leach\footnote{permanent address: School of
Mathematical Sciences, Westville Campus, University of
KwaZulu-Natal, Durban 4000, Republic of South Africa}}
\date{Dipartimento di Matematica e Informatica,
Universit\`a di Perugia, 06123 Perugia, Italy}
\begin {document}

 \maketitle
\begin{abstract}
Searching for a Lagrangian may seem either a trivial endeavour  or
an impossible task. In this paper we show that the Jacobi last
multiplier associated with the Lie symmetries admitted by simple
 models of classical mechanics produces (too?) many Lagrangians in a
simple way. We exemplify the method by such a classic as the
simple harmonic oscillator,  the harmonic oscillator in disguise
[H Goldstein, {\it Classical Mechanics}, 2nd edition
(Addison-Wesley, Reading, 1980)] and the damped harmonic
oscillator. This is the first paper in a series dedicated to this
subject.
\end{abstract}

\section {Introduction}

The last multiplier of Jacobi \cite{Jacobi 42, Jacobi 44 a, Jacobi
44 b, Jacobi 45 a, Jacobi 86 a} is probably nowadays the generally
most forgotten of Jacobi's contributions to Mathematics. Even
after Lie \cite{Lie 74 a} showed that his newly introduced
symmetries provided a very direct route to the calculation of the
multiplier, its use in practice was slight  despite excellent
descriptions of its properties and usage in such classics as the
text of Bianchi \cite{Bianchi 18 a}. For a listing of the
applications of the Jacobi Last Multiplier see the bibliography of
\cite {Nucci 05 b}.  Although Lie groups became of central
importance in some areas of Theoretical Physics, the primary idea
of using infinitesimal transformations to elucidate the properties
of differential equations fell into disuse apart from the
bowdlerised variation known as Buckingham's Theorem which is
widely appreciated by engineers.  With the decline in interest in
the Lie algebraic properties of differential equations the last
multiplier became a mathematical oddity known only to a select
few.

The revival of interest in the Lie algebraic analysis of
differential equations began some fifty years ago, but its
widespread use as a standard tool is only from about half that
period ago.  One of the obstacles to the use of the techniques of
symmetry analysis has always been the high effort in the
computation of the symmetries.  Their calculation could be
characterised as being tedious  without the reward of substantial
intellectual stimulation.  The advent of symbolic manipulation
codes \cite{Hereman 94 a, Hereman 98 a} has allowed the tedium to
be transferred to the computer and freed the brain of the operator
for thinking.

A similar problem of computation besets the calculation of the
Jacobi Last Multiplier. Unfortunately there seems not to have been
the same incentive to use the multiplier as there was in the
application of symmetry methods to the equations of gas dynamics
and it is only in the last decade that Jacobi's Last Multiplier
has seen application in its rightful place with the beneficial
assistance of the computer \cite{Nucci 02 a, Nucci 04 a, Nucci 05
a, Nucci 05 b, Nucci 06 a, Nucci 07 a}.

Jacobi's Last Multiplier is a solution of the linear partial
differential
 equation \cite {Jacobi 44 b, Jacobi 45 a, Jacobi 86 a, Whittaker 44 a},
\begin {equation}
\frac {\partial M} {\partial t} +\sum_{i = 1} ^N \frac {\partial
(Ma_i)} {\partial x_i} = 0, \label{1.0}
\end {equation}
where $\partial_t+\sum_{i = 1} ^N a_i\partial_{x_i} $ is the
vector field of the set of first-order ordinary differential
equations for the $N $ dependent variables $x_i $.  The ratio of
any two multipliers is a first integral of the system of
first-order differential equations and in the case that this
system is derived from the Lagrangian of a one-degree-of-freedom
system one has that \cite {Jacobi 86 a, Whittaker 44 a}
\begin{equation}
\frac{\partial^2L}{\partial \dot{q}^2} = M. \label{1.2}
\end{equation}
Consequently a knowledge of the multipliers of a system enables
one to construct a number of Lagrangians of that system.

We recall that Lie's method \cite {Lie 74 a,Lie 67 a} for the
calculation of the Jacobi Last Multiplier is firstly to find the
value of
\begin{equation}
\Delta = \mbox{\rm det}\left[\begin{array}{c} e_{ij}\\ s_{ij}
\end{array} \right], \label{1.1}
\end{equation}
in which the matrix is square with the elements $e_{ij}$ being the
vector field of the set of first-order differential equations by
which the system is described and the elements, $s_{ij}$, being
the coefficient functions of the number of symmetries of the given
system necessary to make the matrix square. If $\Delta$ is not
zero, the corresponding multiplier is $M=\Delta^{-1}$. Here we
consider that the vector fields of the system of equations and
symmetries are known and that we seek the multiplier.  From
another direction one could know the multiplier and all but one of
the symmetries.  From (\ref{1.1}) the remaining symmetry may be
determined \cite{Nucci 02 a}. Moreover one can use equation
(\ref{1.0}) to raise the order of the system and find nonlocal
symmetries of the original equation \cite{Nucci 05 b}.

When one has a Lagrangian, it is natural to consider the Noether
symmetries of its Action Integral. These constitute a subset, not
necessarily proper, of the Lie symmetries of the corresponding
Lagrangian equation of motion.  The Lagrangians obtained may be
equivalent or inequivalent. The distinction can be variously
expressed.  We use the Noether point symmetries.  If two
Lagrangians have the same number and algebra of Noether point
symmetries, they are equivalent.  Otherwise they are inequivalent.

If we have different Lagrangians, be they equivalent or
inequivalent, it is natural to consider the relationships among
the Lagrangians, the symmetries which produce them through the
Jacobi Last Multiplier, the Noether symmetries of the Lagrangians
and the associated Noether integrals.  This last group returns us
to the Jacobi Last Multipliers since the integrals are ratios of
multipliers.

In this paper we explore the connections mentioned above. For the
purposes of our investigation we choose some simple
one-degree-of-freedom problems with much symmetry.  These are the
simple harmonic oscillator with equation of motion
\begin {equation}
\ddot {q} + k ^ 2q = 0, \label {1.01}
\end {equation}
the simple harmonic oscillator in disguise \cite {Goldstein 80 a}
[ex 18, p 433] with equation of motion
\begin {equation}
q\ddot {q} - 2\dot {q} ^ 2 - k ^ 2q ^ 2 = 0 \label {1.02}
\end {equation}
and the damped linear oscillator with equation of motion
\begin{equation}
\ddot{q} + 2c^2\dot{q} + (c^2+k^2)q = 0, \label{1.3}
\end{equation}
where $c$ and $k$ are constants and in the case of the damped
oscillator the coefficient of $q$ has been written as such to
simplify later expressions. We have made our selection for several
reasons.  Firstly the simple harmonic oscillator is `too good' in
the sense that one of its multipliers is a constant and so all
multipliers are first integrals. This is not the case with (\ref
{1.02}) and (\ref {1.3}).  Since the three equations are related
by point transformations, we do not have to present all of the
details for each equation. In fact under a point transformation
the Lie point symmetries of (\ref {1.01}) become the corresponding
symmetries of (\ref {1.02})/(\ref{1.3}).  The corresponding
multipliers are related according to \cite{Jacobi 44 b, Jacobi 45
a, Jacobi 86 a}
\begin {equation}
\tilde {M} = M |J |, \label {2.6}
\end {equation}
where $J $ is the Jacobian of the transformation.

The paper is structured thusly. In Section 2 we present the Lie
point symmetries of (\ref{1.01}), the associated multipliers and
Lagrangians.  In Section 3 we give the Noether point symmetries of
these Lagrangians and their associated integrals.  The
calculations are routine and so in both sections we simply state
the results.  In \S 4 we present an analysis of these results.

\section{Lie symmetries, Jacobi Last Multipliers and Lagrangians}

To determine Jacobi's Last Multipliers one writes the system under
consideration as a set of first-order ordinary differential
equations.  Thus (\ref{1.01}) becomes
\begin {eqnarray}
& &\dot {u_1} = u_2 \nonumber \\
& &\dot {u_2} = -k ^ 2u_1 \label {10.1}
\end {eqnarray}
with associated vector field
\begin {equation}
X_{SHO} = \partial_t+ u_2\partial_{u_1} - k ^ 2u_1\partial_{u_2}.
\label {10.2}
\end {equation}
We immediately derive that a Jacobi's last multiplier of
(\ref{10.1})/(\ref{10.2}) is a constant, namely 1. In fact
(\ref{1.0}) yields
\begin{equation}\frac{{\rm d}\log M}{{\rm d}t}=-\left[\frac{\partial u_2} {\partial u_1}
+\frac{\partial (-k^2 u_1)} {\partial u_2} \right]=
0.\end{equation} As a linear second-order ordinary differential
equation (\ref {1.01}) possesses eight Lie point symmetries
\cite{Lie 67 a}.  In terms of the variables used in (\ref {10.1})
the eight vectors are
\begin {eqnarray}
\Gamma _{1} &=&  \cos kt{\partial_{u_1}} - k\sin kt{\partial_{u_2}} \nonumber \\
\Gamma _{2} &=& \sin kt{\partial_{u_1}} + k\cos kt{\partial_{u_2}} \nonumber \\
\Gamma _{3} &=& {u_1}{\partial_{u_1}} + {u_2}{\partial_{u_2}} \nonumber \\
\Gamma _{4} &=& \partial_t \nonumber \\
\Gamma _{5} &=& \cos 2kt\partial_t - k{u_1}\sin 2kt{\partial_{u_1}} - \left(2k^{2}{u_1}\cos 2kt  - k{u_2}\sin 2kt\right){\partial_{u_2}} \nonumber \\
\Gamma _{6} &=& \sin 2kt\partial_t + ku_{1}\cos 2kt{\partial_{u_1}} - \left(2k^{2}{u_1}\sin 2kt + k{u_2}\cos 2kt\right){\partial_{u_2}} \nonumber \\
\Gamma _{7} &=& {u_1}\cos kt\partial_t -k{u_1}^{2}\sin kt{\partial_{u_1}} - \left(k^{2}{u_1}^{2}\cos kt + k{u_1}{u_2}\cos kt + {u_2}^{2}\cos kt\right){\partial_{u_2}} \nonumber \\
\Gamma _{8} &=& {u_1}\sin kt\partial_t +k{u_1}^{2}\cos
kt{\partial_{u_1}} - \left(k^{2}{u_1}^{2}\cos kt - k{u_1}{u_2}
\cos kt + {u_2}^{2}\sin kt\right){\partial_{u_2}} \label {10.3}
\end {eqnarray}
in which we have used the ordering: solution symmetries
($\Gamma_1,\Gamma_2$), homogeneity symmetry ($\Gamma_3$),
$sl(2,R)$ symmetries ($\Gamma_3,\Gamma_5,\Gamma_6$) and adjoint
symmetries ($\Gamma_7,\Gamma_8$).  The last are so called since
the time-dependent components of the coefficient functions are
solutions of the equation adjoint to (\ref{1.01}), which in this
case is itself.  For a general second-order equation this be not
the case.

The possible Jacobi Last Multipliers are the reciprocals of the
nonzero determinants
 of the possible matrices of (\ref {1.1}).  We denote them by
$JLM_{ij} $, where the $i $ and $j $ refer to the symmetries used
in the determinants.
  For example $JLM_{12} $ is obtained from
\begin {equation}
\Delta_{12} = \mbox {\rm det}\left [\begin {array} {ccc}
1 &u_2 & -k ^ 2u_1\\
0&\cos kt & - k\sin kt \\
0& \sin kt &  k\cos kt
\end {array}\right]
= k. \label {10.4}
\end {equation}
In the matrix the second row comes from $\Gamma_1 $ and the third
from $\Gamma_2 $.
 Then we obtain
\begin {eqnarray}
& & JLM_{12} = k \nonumber \\
& & JLM_{13} = \frac{1}{ku_1\sin k t+u_2\cos kt} \nonumber \\
& & JLM_{14} = -\frac{1}{k}\,JLM_{23}\nonumber \\
& & JLM_{15} = \frac{1}{k}\,JLM_{23} \nonumber \\
& & JLM_{16} = -\frac{1}{k}\,JLM_{13} \nonumber \\
& & JLM_{17} = -JLM_{13}^2 =-\frac{1}{(ku_1\sin k t+u_2\cos kt)^2} \nonumber \\
& & JLM_{18} = -JLM_{13}\times JLM_{23}=\frac{1}{(ku_1\sin k t+u_2\cos kt)(ku_1\cos k t-u_2\sin kt)}  \nonumber \\
& & JLM_{23} = \frac{1}{-ku_1\cos k t+u_2\sin kt} \nonumber \\
& & JLM_{24} = \frac{1}{k}\,JLM_{13} \nonumber \\
& & JLM_{25} = \frac{1}{k}\,JLM_{13} \nonumber \\
& & JLM_{26} = \frac{1}{k}\,JLM_{23} \nonumber \\
& & JLM_{27} = JLM_{18} \nonumber \\
& & JLM_{28} = -JLM_{23}^2 =-\frac{1}{(-ku_1\cos k t+u_2\sin
kt)^2}\nonumber\\
& & JLM_{34} = \left [JLM_{13}^{-2} + JLM_{23}^{-2}\right] ^ {- 1} =\frac{1}{u_2^2+k^2u_1^2}\nonumber \\
& & JLM_{35} = \left [JLM_{13}^{-2} - JLM_{23}^{-2}\right] ^ {- 1}=\frac{1}{[(u_2^2-k^2u_1^2)\cos 2kt+2ku_1u_2\sin 2kt]} \nonumber \\
& & JLM_{36} = -\frac{1}{2}JLM_{18}\nonumber \\
& & JLM_{37} = 0\nonumber \\
& & JLM_{38} = 0\nonumber \\
& & JLM_{45} = \frac{1}{2k}\,JLM_{18} \nonumber \\
& & JLM_{46} = \frac{1}{k}\,JLM_{35} \nonumber \\
& & JLM_{47} = JLM_{13} JLM_{34}=\frac{1}{(u_2^2+k^2u_1^2)(ku_1\sin k t+u_2\cos kt)} \nonumber \\
& & JLM_{48} = JLM_{23} JLM_{34}=\frac{1}{(u_2^2+k^2u_1^2)(-ku_1\cos k t+u_2\sin kt)} \nonumber \\
& & JLM_{56} = \frac{1}{k}\,JLM_{34} \nonumber \\
& & JLM_{57} = JLM_{13} JLM_{35}=\frac{1}{(ku_1\sin k t+u_2\cos kt)[(u_2^2-k^2u_1^2)\cos 2kt+2ku_1u_2\sin 2kt]} \nonumber \\
& & JLM_{58} = JLM_{23} JLM_{35}=\frac{1}{(-ku_1\cos k t+u_2\sin kt)[(u_2^2-k^2u_1^2)\cos 2kt+2ku_1u_2\sin 2kt]} \nonumber \\
& & JLM_{67} =\frac{1}{2}\, JLM_{13}^2 JLM_{23}=\frac{1}{2(ku_1\sin k t+u_2\cos kt)^2(-ku_1\cos k t+u_2\sin kt)} \nonumber \\
& & JLM_{68} = \frac{1}{2}\, JLM_{13} JLM_{23}^2=\frac{1}{2(ku_1\sin k t+u_2\cos kt)(-ku_1\cos k t+u_2\sin kt)^2}\nonumber \\
& & JLM_{78} = 0. \label {10.5}
\end {eqnarray}
Note that there are 14 different multipliers (multiplicative
constants are inessential).

For each of these 14 multipliers we can calculate a Lagrangian
using (\ref {1.2}).
  One readily appreciates that two arbitrary functions of
integration in the variables $t $ and $u_1 $ are introduced with
the double quadrature, say $f_1(t,u_1)$ and $f_2(t,u_1)$.  This
means that there is a doubly infinite family of Lagrangians
corresponding to each multiplier. There must be a constraint on
these functions for the application of the variational principle
to give precisely (\ref {1.01}). This constraint is listed with
the Lagrangians.  We use the same subscripts to identify the
Lagrangians.
  The constraint is placed after the Lagrangian to which it applies.
 \begin {eqnarray}
 & &L_{12} = \half u_{2}^2+f_1u_{2}+f_2, \label {10.6} \\
& &\nonumber\\
& &\frac {\partial f_1}{\partial t} - \frac {\partial f_2}{\partial u_{1}} =  k^2u_{1}; \nonumber \\
& &\nonumber\end{eqnarray}
\begin {eqnarray} & &\hspace{-0.6cm}L_{13} =
\sec^2kt\left[\log (ku_{1}\sin kt  + u_{2}\cos kt
)\left(ku_{1}\sin kt
  + u_{2}\cos kt \right)
 -u_{2}\cos kt  - ku_{1}\sin kt\right] +f_1u_{2}+f_2, \label {10.9} \\
& &\nonumber\\
& & {\frac {\partial f_1}{\partial t}} - {\frac {\partial f_2}{\partial u_{1}}} = 0; \nonumber \\
& &\nonumber\end{eqnarray}
\begin {eqnarray} & &\hspace{-1cm}L_{23} =
\cosec^2kt\left[\log (-ku_{1}\cos kt + u_{2}\sin kt
)\left(-ku_{1}\cos kt
 +u_{2}\sin kt \right) -u_{2}\sin kt  + ku_{1}\cos kt \right]+f_1u_{2}+f_2, \label {10.7} \\
& &\nonumber\\
& & {\frac {\partial f_1}{\partial t}} - {\frac {\partial f_2}{\partial u_{1}}} = 0; \nonumber \\
& &\nonumber\end{eqnarray}
\begin {eqnarray} & & L_{17} = -\sec^2
kt \log \left (ku_{1}\sin kt +u_{2}\cos kt \right )
+f_1u_{2}+f_2 ,\label {10.8} \\
& &\nonumber\\
& &\frac {\partial f_1} {\partial t} -\frac {\partial f_2} {\partial u_{1}} = 0; \nonumber \\
& &\nonumber\end{eqnarray}
\begin {eqnarray} & & L_{18} =
\frac{1}{ku_{1}\sin kt \cos kt } \Big[\sin kt (ku_{1}\sin kt +
u_{2}\cos kt)\log\left(ku_{1}\sin kt + u_{2}\cos kt\right)
\nonumber \\&&\qquad \;\;+\cos kt(-u_{2}\sin kt+ku_{1}) \log
\left(-u_{2}\sin kt+ku_{1} \right)\Big]
+f_1u_{2}+f_2, \label {10.10} \\
& &\nonumber\\
& &\left( u_{1}\sin  kt\cos kt \right) \left(\frac{\partial
f_1}{\partial t}- \frac{\partial f_2}{\partial u_{1}}\right) =
 1; \nonumber \\
& &\nonumber\end{eqnarray}\begin {eqnarray}
& &L_{28} = \cosec^2kt\log (ku_{1}\cos kt -u_{2}\sin kt )+f_1u_{2}+f_2, \label {10.11} \\
& &\nonumber\\
& &\frac {\partial f_1}{\partial t} - \frac {\partial f_2}{\partial u_{1}} = 0; \nonumber \\
& &\nonumber\end{eqnarray}\begin{eqnarray} && L_{34}=
\frac{u_2}{ku_1}\,\arctan\left(\frac{u_2}{ku_1}\right)-\frac{1}{2}\,\log\left(\frac{u_2^2}{k^2u_1^2}+1\right)+f_1u_2+f_2,
\label
{10.11b}\\& &\nonumber\\
& & u_{1} \left(\frac{\partial f_1}{\partial t}- \frac{\partial
f_2}{\partial u_{1}}\right) =
 1; \nonumber \\
& &\nonumber
\end{eqnarray}
\begin {eqnarray} & & L_{35} =
{\displaystyle \frac {1}{2ku_{1}\cos 2kt}} \Big[2ku_1+
\left({u_{2}}{\cos}2 k t  + ku_{1}\sin 2kt-ku_1\right)\,
 \log\Big((\sin kt+\cos kt)u_2+(\sin kt-\cos kt)ku_1\Big) \nonumber \\
& &\qquad \;-\left({u_{2}}{\cos}2 kt+ ku_{1}\sin 2kt+ku_1\right)\,
\log\Big((\sin kt-\cos kt)u_2-(\sin kt+\cos kt)ku_1\Big)\Big]
+ {f_1}{u_{2}} + {f_2},\label {10.12} \\
& &\nonumber\\
& & u_{1}\cos 2kt\left({\frac {\partial f_1}{\partial t}}
  - {\frac {\partial f_2}{\partial u_{1}}}\right) = 1; \nonumber \\
& &\nonumber\end{eqnarray}\begin {eqnarray} & &L_{47} = \frac{1}{2
k^2u_1^2}\left[-(u_2\cos k t+k u_1 \sin k t) \log
(u_2^2+k^2u_1^2)+2(u_2\sin k t- ku_1\cos k t
)\arctan\left(\frac{u_2}{ku_1}\right)\right .\nonumber\\
&&\qquad +2(u_2\cos k t +k u_1 \sin k t)\log(u_2\cos k t+k
u_1 \sin k t)\Big]+{f_1}{u_{2}} + {f_2},\label {10.13} \\
& &\nonumber\\
& & k u_1^2\left(\frac {\partial f_1}{\partial t}- \frac {\partial
f_2}{\partial u_{1}} \right)= \sin k t ;\nonumber \\
& &\nonumber\end{eqnarray}\begin {eqnarray} & &L_{48} =
\displaystyle {\frac {1}{2 k^2 u_1^2}} \left[-(u_2\sin k t-
ku_1\cos k t)\log(u_2^2+k^2u_1^2) -2(u_2\cos k t +ku_1\sin k t)
\arctan\left(\frac{u_2}{k u_1}\right)\right .\nonumber\\
&&\qquad +2(u_2\sin k t- ku_1\cos k t)\log(u_2\sin k t- ku_1\cos k
t)\Big]+ {f_1}{u_{2}}  + {f_2},\label {10.14} \\
& &\nonumber\\
& &  k u_1^2\left( {\frac {\partial f_1}{\partial t}}
  - {\frac
{\partial f_2}{\partial u_{1}}} \right) = -\cos k t;\nonumber \\
& &\nonumber\end{eqnarray}\begin {eqnarray} & & L_{57} =
\frac{1}{2k^2u_1^2}\Big[ \Big((\cos k t -\sin k t)u_2+(\cos k t
+\sin k t)ku_1 \Big)\,\log\Big((\cos k t -\sin k t)u_2+(\cos k t
+\sin k t)ku_1 \Big)\nonumber\\
&&\qquad + \Big((\cos k t +\sin k t)u_2-(\cos k t -\sin k t)ku_1
\Big)\,\log\Big((\cos k t +\sin k t)u_2-(\cos k t
-\sin k t)ku_1 \Big)\nonumber\\
&&\qquad -2(u_2\cos k t+ku_1\sin k t)\,\log(u_2\cos k t+ku_1\sin k t)\Big]+f_1u_{2}+f_2, \label {10.15} \\
& &\nonumber\\
& & \frac {\partial f_1} {\partial t} -\frac {\partial f_2}
{\partial u_{1}} = 0;
 \nonumber \\
& &\nonumber\end{eqnarray}\begin {eqnarray} & &L_{58} =
\frac{1}{2k^2u_1^2}\Big[ \Big((\cos k t -\sin k t)u_2+(\cos k t
+\sin k t)ku_1 \Big)\,\log\Big((\cos k t -\sin k t)u_2+(\cos k t
+\sin k t)ku_1 \Big)\nonumber\\
&&\qquad - \Big((\cos k t +\sin k t)u_2-(\cos k t -\sin k t)ku_1
\Big)\,\log\Big((\cos k t +\sin k t)u_2-(\cos k t
-\sin k t)ku_1 \Big)\nonumber\\
&&\qquad +2(u_2\sin k t-ku_1\cos k t)\,\log(u_2\sin k t-ku_1\cos k
t)\Big]
+f_1u_{2}+f_2, \label {10.16} \\
& &\nonumber\\
& & {\frac {\partial f_1}{\partial t}}
  - {\frac {\partial f_2}{\partial u_{1}}} =0;
 \nonumber \\
& &\nonumber\end{eqnarray}\begin {eqnarray} & &L_{67} = \frac{k
u_1\cos k t -u_2\sin k t}{2k^2u_1^2}\,\Big[\log(k u_1\sin k t
+u_2\cos k t)   -\log(u_2\sin k t -k u_1\cos k t) \Big] +
{f_1}{u_{2}}
 + {f_2}, \label {10.17}\\
& &\nonumber\\
& & {\frac {\partial f_1}{\partial t}}
  - {\frac {\partial f_2}{\partial u_{1}}} =0;
  \nonumber \\
& &\nonumber
\end {eqnarray}
\begin {eqnarray} & &L_{68} =
\frac{k u_1\sin k t +u_2\cos k t}{2k^2u_1^2}\,\Big[\log(k u_1\sin
k t +u_2\cos k t)   -\log(u_2\sin k t -ku_1\cos k t)\Big]  +
{f_1}{u_{2}}
 + {f_2}, \label {10.18}\\
& &\nonumber\\
& & {\frac {\partial f_1}{\partial t}}
  - {\frac {\partial f_2}{\partial u_{1}}} =0.
\nonumber \\
& &\nonumber
\end{eqnarray}

We illustrate the importance of the two functions $f_1$ and $f_2$
by a consideration of $L_{12}$.  The constraint is
$$
\frac {\partial f_{1}} {\partial t} -\frac {\partial f_{2}}
{\partial u_1} = k^2u_1
$$
which is satisfied if we write
\begin {equation}
f_1 = \frac {\partial g} {\partial u_1},\quad f_2 = \frac
{\partial g} {\partial t} - k^2u_1, \nonumber
\end {equation}
where $g (t,u_1) $ is an arbitrary function of its arguments.
Hence
\begin {equation}
L_{12} = \half\left (u_2^2 - k^2u_1 ^2\right) + \frac{\d g}{\d t}.
\label {1.8}
\end {equation}

We see that for $L_{12}$ the two arbitrary functions introduced on
the integration of the Jacobi Last Multiplier to obtain the
Lagrangian are subject to a constraint to obtain compatibility
with the Newtonian equation of motion under consideration.  The
remaining generality can be expressed as the total time derivative
of an arbitrary gauge function which has no effect upon the form
of the Lagrangian equation of motion. There are some further
considerations.  The function, $g(t,u_1)$, is by its very
construction a function only of the indicated variables.
Consequently one would expect $L_{12}$ to have the same number of
Noetherian point symmetries for all possible functions,
 $g(t,u_1)$, since it may be absorbed into the boundary
function of Noether's Theorem \cite{Noether 18 a}. This conflation
with the boundary term may
 be part of the explanation for some of the confused accounts of
Noether's Theorem which one finds in the literature from some
twenty years \cite {Courant 40 a} after Noether's seminal work
until the almost present day \cite {Dresner 98 a}.  Again we used
$L_{12}$ to demonstrate the result, but we expect that the
informed reader appreciates that this is general.

In the case of (\ref {1.02}) the corresponding system of
first-order equations is
\begin {eqnarray}
\dot u_1 & = & u_2 \nonumber \\
\dot u_2 & = & \displaystyle {\frac {1} {u_1}}\left (2u_2 ^ 2+u_1
^ 2\right) \label {2.8}
\end {eqnarray}
with associated vector field
\begin {equation}
X_{GHO} = \partial_t+ u_2\partial_{u_1} + \frac {1} {u_1}\left
(2u_2 ^ 2+u_1 ^ 2\right)\partial_{u_2}. \label {2.8bis}
\end {equation}
It is easy to derive \cite{Jacobi 45 a}[\S 18] that a Jacobi's
last multiplier of (\ref{2.8})/(\ref{2.8bis}) is $u_1^{-4}$. In
fact (\ref{1.0}) yields
\begin{equation}\frac{{\rm d}\log M}{{\rm
d}t}=-\left[\frac{\partial u_2} {\partial u_1} +\frac{\partial }
{\partial u_2}  \left (\frac{2u_2 ^ 2+u_1 ^ 2}{u_1}\right)\right]=
-\frac{4u_2}{u_1}.\end{equation}
  The eight Lie point symmetries are
\begin {eqnarray}
\Gamma_1 & = & u_1 ^ 2\cos t\partial_{u_1} - \left (u_1\sin t- 2u_1\cos t\right)\partial_{u_2} \nonumber\\
\Gamma_2 & = & u_1 ^ 2\sin t\partial_{u_1} +\left (u_1\cos t + 2u_1\sin t\right)\partial_{u_2} \nonumber\\
\Gamma_3 & = & u_1\partial_{u_1} + u_2\partial_{u_2} \nonumber\\
\Gamma_4 & = & \partial_t \nonumber\\
\Gamma_5 & = & \cos 2t\partial_t+ u_1\sin 2t\partial_{u_1} +\left (2u_1\cos 2t+ 3u_2\sin 2t\right)\partial_{u_2} \nonumber\\
\Gamma_6 & = & \sin 2t\partial_t - u_1\cos 2t\partial_{u_1} +\left (2u_1\sin 2t - 3u_2\cos 2t\right)\partial_{u_2} \nonumber\\
\Gamma_7 & = & \displaystyle {\frac {1} {u_1 ^ 2}}\left [u_1\cos t\partial_t+ u_1 ^ 2\sin t\partial_{u_1} + \left (u_1 ^ 2\cos t +u_1u_2\sin t + u_2\cos t\right)\partial_{u_2}\right] \nonumber\\
\Gamma_8 & = & \displaystyle {\frac {1} {u_1 ^ 2}}\left [u_1\sin
t\partial_t- u_1 ^ 2\cos t\partial_{u_1} + \left (u_1 ^ 2\sin t
-u_1u_2\cos t + u_2\sin t\right)\partial_{u_2}\right]. \label
{2.9}
\end {eqnarray}
Not surprisingly there are fourteen different
 Jacobi Last Multipliers, i.e.:
\begin {eqnarray}
JLM_{12} & = & \displaystyle {\frac {1} {u_1 ^ 4}} \nonumber\\
JLM_{13} & = & \displaystyle {\frac {1} {u_1 ^ 2 (u_1\sin t -u_2\cos t)}} \nonumber\\
JLM_{23} & = & \displaystyle {\frac {1} {u_1 ^ 2 (u_1\cos t +u_2\sin t)}} \nonumber\\
JLM_{17} & = & \displaystyle {\frac {1} {(u_1\sin t-u_2\cos t) ^ 2}} \nonumber\\
JLM_{18} & = & \displaystyle {\frac {1} {(u_1\sin t-u_2\cos t) (u_1\cos t+u_2\sin t) }} \nonumber\\
JLM_{28} & = & \displaystyle {\frac {1} {(u_1\cos t +u_2\sin t) ^ 2}} \nonumber\\
JLM_{34} & = & \displaystyle {\frac {1} {u_1 ^ 2+u_2 ^ 2}} \nonumber\\
JLM_{35} & = & \displaystyle {\frac {1} {(u_1 ^ 2-u_2 ^ 2)\cos 2t+ 2u_1u_2\sin 2t}} \nonumber\\
JLM_{47} & = & \displaystyle {\frac {u_1^2} {\left (u_1 ^ 2+u_2 ^ 2\right) (u_1\sin t-u_2\cos t) }} \nonumber\\
JLM_{48} & = & \displaystyle {\frac {u_1^2} {\left (u_1 ^ 2+u_2 ^
2\right) (u_1\cos t+u_2\sin t) }}
 \nonumber\\
JLM_{57} & = & \displaystyle {\frac {u_1 ^ 2}
 {\left ((u_1 ^ 2-u_2 ^ 2)\cos 2t+ 2u_1u_2\sin 2t\right) (u_1\sin t-u_2\cos t) }}
 \nonumber\\
JLM_{58} & = & \displaystyle {\frac {u_1 ^ 2} {\left ((u_1 ^ 2-u_2
^ 2)\cos 2t+ 2u_1u_2\sin 2t\right) (u_1\cos t + u_2\sin t) }}
\nonumber\\ JLM_{67} & = & \displaystyle {\frac {u_1 ^ 2}
 {\left ((u_1 ^ 2-u_2 ^ 2)\sin 2t - 2u_1u_2\cos 2t\right) (u_1\sin t-u_2\cos t) }} \nonumber\\
JLM_{68} & = & \displaystyle {\frac {u_1 ^ 2}
 {\left ((u_1 ^ 2-u_2^ 2)\sin 2t - 2u_1u_2\cos 2t\right) (u_1\cos t +u_2\sin t)
 }}\,.
\label{2.10}
\end{eqnarray}
As we have already provided the details of the Lagrangians in the
case of (\ref {1.01}), we do not repeat the whole listing for
(\ref {1.02}).  However, we do give two Lagrangians and their
associated Hamiltonians.  In the case of $JLM_{12} $ the
Lagrangian is
\begin {equation}
L_{12} = \half\frac {u_2 ^ 2} {u_1 ^ 4} +f_1 u_2+f_2 \label {2.13}
\end {equation}
subject to the constraint
\begin {equation}
\frac {\partial f_1} {\partial t} -\frac {\partial f_2} {\partial
u_1} = -\frac {1} {u_1 ^ 3}. \label {2.14}
\end {equation}
It is a simple matter to show that the corresponding Hamiltonian
is
\begin {equation}
H_{12} = \half q ^ 4\left (p -f_1\right) ^ 2-f_2, \label {2.12}
\end {equation}
where we have used the traditional symbols for the canonical
variables with $q =u_1 $ and $p =\partial L_{12}/\partial u_2 $. A
particular solution of (\ref{2.14}) is $f_1=0, f_2=1/(2u_1^2)$,
which substituted into (\ref{2.12}) yields the Hamiltonian in
\cite {Goldstein 80 a}, i.e.:
\begin {equation}
H_{G} = \half \left( q ^ 4 p ^ 2+\frac{1}{q^2}\right). \label
{2.12b}
\end {equation}
 In the case of $JLM_{34} $ the
Lagrangian is
\begin {equation}
L_{34} = \frac {u_2} {u_1}\arctan\left (\frac {u_2} {u_1}\right) -
\half\log\left (\left (\frac {u_2} {u_1}\right) ^ 2+ 1\right) +f_1
u_2+f_2  \label {2.11}
\end {equation}
subject to the constraint
\begin {equation}
\frac {\partial f_1} {\partial t} -\frac {\partial f_2} {\partial
u_1} = -\frac {1} {u_1}.  \label {2.110}
\end {equation}
The corresponding Hamiltonian is
\begin {equation}
H_{34} = \half \log\left (\tan(q(p-f_1))^2+1\right) -f_2. \label
{2.112}
\end {equation}
 In general, due to the presence of
the arbitrary function in the Hamiltonian containing the
independent variable, the Hamiltonian is not a constant of the
motion.  However, the ratio of the multipliers is a first
integral.  For example we have that both
\begin {equation}
\frac {JLM_{13}} {JLM_{47}} = \frac {u_2 ^ 2+u_1 ^ 2} {u_1 ^ 4}
\label {2.15}
\end {equation}
and
\begin {equation}
\frac {JLM_{23}} {JLM_{58}} = \frac {(u_1 ^ 2-u_2 ^ 2)\cos 2t+
2u_1u_2\sin 2t} {u_1 ^ 4} \label {2.15b}
\end {equation}
are first integrals of (\ref{2.8}).

In the case of (\ref {1.3}) the associated system of first-order
equations is
\begin{eqnarray}
&&\dot{u_1} = u_2 \nonumber \\
&&\dot{u_2} = -\left(c^2+k^2\right)u_1-2cu_2 \label{2.1}
\end{eqnarray}
with associated vector field
\begin{equation}
X = \partial_t +u_2\partial_{u_1} -
\left[\left(c^2+k^2\right)u_1+2cu_2\right]\partial_{u_2}.
\label{2.2}
\end{equation}
It is easy to derive \cite{Jacobi 86 a} that a Jacobi's last
multiplier of (\ref{2.1})/(\ref{2.2}) is $\exp[2ct]$. In fact
(\ref{1.0}) yields
\begin{equation}\frac{{\rm d}\log M}{{\rm d}t}=-\frac{\partial u_2} {\partial u_1}
+\frac{\partial } {\partial
u_2}\left[\left(c^2+k^2\right)u_1+2cu_2\right] = 2c.\end{equation}
 As a linear second-order equation the
damped linear oscillator described by (\ref{1.3}) possesses eight
Lie point symmetries \cite{Lie 67 a} which have been given
explicitly in \cite{Leach 78 c}. In terms of the variables
introduced in (\ref{2.1}) the eight vectors are
\begin{eqnarray}
&&\Gamma_1 = \exp[-ct]\left[\cos kt \partial_{u_1} - (c\cos kt+k\sin kt)\partial_{u_2}\right] \nonumber \\
&&\Gamma_2 = \exp[-ct]\left[-\sin kt \partial_{u_1} + (c\sin
kt-k\cos kt)\partial_{u_2}\right] \nonumber \\
&& \Gamma_3 = u_1\partial_{u_1}+u_2\partial_{u_2} \nonumber \\
&& \Gamma_4 = \partial_t \nonumber \\
&& \Gamma_5 = \cos 2kt\partial_t -u_1\left(c\cos 2kt + k \sin 2kt\right)\partial_{u_1} \nonumber \\
&&\qquad +\left[2ku_1\left(c\sin 2kt -k\cos 2kt\right)
-u_2(c\cos 2kt-k\sin 2kt) \right]\partial_{u_2}  \nonumber \\
&& \Gamma_6 = \sin 2kt\partial_t - u_1\left(c\sin 2kt - k\cos 2kt\right)\partial_{u_1} \nonumber \\
&&\qquad+\left[-2ku_1(k\sin 2kt + c\cos 2kt)
- u_2(c\sin 2kt + k\cos 2kt)\right]\partial_{u_2} \nonumber \\
&& \Gamma_7 = \exp[ct]\left[u\cos kt\partial_t - u_1^2\left(c\cos
kt +k\sin kt \right)\partial_{u_1} \right. \nonumber \\
&&\qquad \left.- \left(u_2^2\cos kt+u_1u_2(3c\cos kt + k\sin kt)+
(c^2+k^2)u_1^2\cos kt\right)\partial_{u_2}\right]  \nonumber \\
&& \Gamma_8 = \exp[ct]\left[-u_1\sin kt\partial_t +
u_1^2\left(c\sin kt -k\cos kt \right)\partial_{u_1}  \right. \nonumber \\
&&\qquad\left. +\left(u_2^2\sin kt-u_1u_2(k\cos kt - 3c \sin kt)+
(c^2+k^2)u_1^2\sin kt\right)\partial_{u_1}\right].
 \label{2.3}
\end{eqnarray}

We calculate the Jacobi Last Multipliers as we did for (\ref
{1.01}).  Of course there are fourteen different
 Jacobi Last Multipliers, namely:
\begin {eqnarray}
& &JLM_{12} = \exp[2ct]\nonumber\\
&& JLM_{13} = \displaystyle{\frac {\exp[ct]}{ u_1(c\cos kt +k\sin
kt)+ u_2\cos kt }}
\nonumber \\
&& JLM_{23} = \displaystyle{\frac {\exp[ct]}{  u_1(c\sin kt - k\cos kt) + u_2\sin kt}} \nonumber \\
&& JLM_{17} = \displaystyle{\frac{1}{ \left[u_1(c\cos kt+ k\sin kt) +u_2\cos kt\right]^2 }}\nonumber \\
&& JLM_{18} = \displaystyle{\frac{1}{\left[u_1(c\cos kt+ k\sin kt) +u_2\cos kt\right]\left[u_1(c\sin kt - k\cos kt) +u_2\sin kt\right]}} \nonumber \\
&& JLM_{28} = \displaystyle{\frac{1}{\left[ u_1(c\sin kt - k\cos kt)+{u_2}\sin kt\right] ^{2}}} \nonumber \\
&& JLM_{34} = \displaystyle{\frac{1}{(cu_1+u_2)^2 + k^2u_1^2}} \nonumber \\
&& JLM_{35} = \displaystyle{\frac{1}{ 2ku_1(cu_1+u_2)\sin 2kt+[(cu_1+u_2)^2-k^2u_1^2] \cos 2kt}} \nonumber \\
&& JLM_{47} = \displaystyle{\frac {1}{{\exp[ct]}
 \left[ (cu_1+u_2)^2 + k^2u_1^2\right]
  \left[u_1(c\cos kt +k\sin kt)+ u_2\cos kt\right]}} \nonumber \\
&& JLM_{48} = \displaystyle{\frac {1}{{\exp[ct]} \left[
(cu_1+u_2)^2 + k^2u_1^2\right]
 \left[  u_1(c\sin kt - k\cos kt) + u_2\sin kt \right]}} \nonumber \\
&& JLM_{57} =
\displaystyle{\frac {1}{{\exp[ct]}
 \left[ u_1(c\cos kt +k\sin kt)+ u_2\cos kt\right]
  \left[2ku_1(cu_1+u_2)\sin 2kt+[(cu_1+u_2)^2-k^2u_1^2] \cos 2kt \right]}} \nonumber
  \\
 && JLM_{58} =
\displaystyle{\frac {1}{{\exp[ct]}
 \left[ u_1(c\sin kt - k\cos kt) + u_2\sin kt\right]
  \left[2ku_1(cu_1+u_2)\sin 2kt+[(cu_1+u_2)^2-k^2u_1^2] \cos 2kt \right]}} \nonumber \\
&& JLM_{67} = \displaystyle{\frac {1}{{\exp[ct]}
 \left[ u_1(c\cos kt +k\sin kt)+ u_2\cos kt \right]^2
  \left[ u_1(c\sin kt - k\cos kt) + u_2\sin kt \right]}} \nonumber
  \\
&& JLM_{68} =
\displaystyle{\frac {1}{{\exp[ct]}
 \left[ u_1(c\cos kt +k\sin kt)+ u_2\cos kt \right]
  \left[ u_1(c\sin kt - k\cos kt) + u_2\sin kt\right]^2}}\,.
 \label {2.5}
\end {eqnarray}
As we have already provided the details of the Lagrangians in the
case of (\ref {1.01}), we do not repeat the whole listing for
(\ref {1.3}).  We show just three Lagrangians, and corresponding
constraint, i.e.:
 \begin {eqnarray}
 & &L_{12} = \half \exp[2ct]u_{2}^2+f_1u_{2}+f_2, \label {12.6}
 \\
& &\nonumber\\
& &\frac {\partial f_1}{\partial t} - \frac {\partial
f_2}{\partial u_{1}} =
 \exp[2ct](c^2+ k^2)u_{1}; \nonumber \\
& &\nonumber\end{eqnarray}
\begin{eqnarray} && L_{34}=
\left(\frac{u_2}{ku_1}+\frac{c}{k}\right)\,\arctan\left(\frac{u_2}{ku_1}+\frac{c}{k}\right)
-\frac{1}{2}\,\log\left(\left(\frac{u_2}{ku_1}+\frac{c}{k}\right)^2+1\right)+f_1u_2+f_2,
\label
{12.3}\\& &\nonumber\\
& & u_{1} \left(\frac{\partial f_1}{\partial t}- \frac{\partial
f_2}{\partial u_{1}}\right) =
 1; \nonumber \\
& &\nonumber
\end{eqnarray}
\begin {eqnarray} & &L_{67} =-\frac{u_2\sin 2kt+(c \sin 2 kt+2k\sin^2 kt)u_1}{2\exp[ct]k^2u_1^2\sin kt}\,
\log\,\left[\displaystyle{\frac{u_2\cos k t+(c\cos kt+k\sin kt)
u_1 }{u_2\sin k t +(c\sin kt -k\cos k t)
 u_1}}\right] + {f_1}{u_{2}}
 + {f_2}, \label {13.17}\\
& &\nonumber\\
& & {\frac {\partial f_1}{\partial t}}
  - {\frac {\partial f_2}{\partial u_{1}}}
  =0
  \nonumber \\
& &\nonumber
\end {eqnarray}

\section {Noether Symmetries and Integrals}

For the purposes of this Section we confine our interest to
Noetherian point symmetries being well aware that Noether did not
such thing \cite {Noether 18 a}.  It is known that the number of
Noetherian point symmetries of the Action Integral for a
Lagrangian of standard form for a one-degree-of-freedom mechanical
system is 0, 1, 3 or 5 \cite {Mahomed 93 a, Kara 94 a}.  We list
the symmetries and integrals in classes according to the number of
Noetherian point symmetries, which turn to be 5, 3, or 2.  Since
the results are the same for each of the equations under
consideration, we give examples only for (\ref
{1.01}) but in the notation of (\ref {10.1}).\\

\noindent{\bf Five Noetherian point symmetries}\newline
$$
\begin {array} {llcl}
L_{12}\quad &\Gamma_1 &\Rightarrow& -u_2\cos kt - ku_1\sin kt \\
[0.2cm] &\Gamma_2 &\Rightarrow& u_2\sin kt - ku_1\cos kt \\
[0.2cm] &\Gamma_4 &\Rightarrow&\half\left (u_2 ^ 2+ k ^ 2 u_1
^2\right) \\ [0.2cm] &\Gamma_5 &\Rightarrow& \half\left (u_2 ^ 2 -
k ^ 2u_1 ^ 2\right)\cos 2kt + ku_1u_2\sin 2kt \\
[0.2cm] &\Gamma_6 &\Rightarrow& -\half\left (u_2 ^ 2 - k ^ 2u_1 ^
2\right)\sin 2kt + ku_1u_2\cos 2kt\\ [0.2cm] \\
\end {array}
$$
{\bf Three Noetherian point symmetries}\newline
$$
\begin {array} {llcl}
L_{13}\quad &\Gamma_1 &\Rightarrow& \log\left (u_2\cos kt +
ku_1\sin kt\right) \\ [0.1cm] &\Gamma_4+\Gamma_5 &\Rightarrow&
2\left(u_2\cos kt +ku_1\sin kt\right) \\
[0.1cm] &-k\Gamma_3+\Gamma_6 &\Rightarrow& 2\left(-ku_1\cos kt +
u_2\sin kt\right) \\ [0.1cm]\\ \hline\\  L_{23}\quad &\Gamma_2
&\Rightarrow& \log\left (-ku_1\cos kt + u_2\sin kt\right) \\
[0.1cm] &-\Gamma_4+\Gamma_5 &\Rightarrow&
2\left(ku_1\cos kt -u_2\sin kt\right) \\
[0.1cm] &k\Gamma_3+\Gamma_6 &\Rightarrow& 2\left(u_2\cos kt +ku_1\sin kt\right) \\ [0.1cm]\\
\hline \\L_{28}\quad &\Gamma_2 &\Rightarrow&
\displaystyle {\frac {1} {ku_1\cos kt - u_2\sin kt}} \\
[0.4cm] &\Gamma_3 &\Rightarrow& \displaystyle{\frac{1}{k}}\,
\displaystyle {\frac { ku_1\sin kt + u_2\cos kt} {ku_1\cos kt -
u_2\sin kt}} \\ [0.4cm]
 &-\Gamma_4+\Gamma_5 &\Rightarrow& 2\left(\log\left
(ku_1\cos kt - u_2\sin kt\right) -1\right)\\  [0.1cm]\\
\hline\end{array}$$
$$\begin{array}{llcl}
L_{67}\quad &k\Gamma_3 +\Gamma_6&\Rightarrow& \displaystyle {\frac
 {-ku_1\cos kt + u_2\sin kt}{-\half\left (u_2^ 2 - k^ 2u_1^2\right)\sin 2kt + ku_1u_2\cos 2kt}}
\\
[0.5cm] &\Gamma_7 &\Rightarrow&
\displaystyle{\frac{1}{2k}}\,\log\left
(\displaystyle{\frac{-ku_1\cos kt + u_2\sin kt}{u_2\cos kt +
ku_1\sin kt}}\right)+\displaystyle{\frac{1}{2k}}
\\ [0.5cm]
 &\Gamma_8 &\Rightarrow& \displaystyle{\frac{1}{k}}\,\displaystyle {\frac{ku_1u_2\sin 2kt-(u_2^ 2 - k^ 2u_1^2)\cos^2 kt-
 u_2^2\sin^2 kt}{-\left (u_2^ 2 - k^ 2u_1^2\right)\sin 2kt + 2ku_1u_2\cos 2kt}} \\  [0.2cm]\\
 \hline\\
L_{68}\quad & -k\Gamma_3 +\Gamma_6 &\Rightarrow& -\displaystyle
{\frac {u_2\cos kt + ku_1\sin kt}{\half\left (u_2^ 2 - k^
2u_1^2\right)\sin 2kt - ku_1u_2\cos 2kt}}
 \\
[0.5cm] &\Gamma_7 &\Rightarrow&
\displaystyle{\frac{1}{k}}\,\displaystyle {\frac{ku_1u_2\sin
2kt+k^ 2u_1^2\sin^2 kt+ u_2^2\cos^2 kt}{-\left (u_2^ 2 - k^
2u_1^2\right)\sin 2kt + 2ku_1u_2\cos2 kt}}
\\ [0.5cm]
 &\Gamma_8 &\Rightarrow& \displaystyle{\frac{1}{2k}}\,\log\left
(\displaystyle{\frac{-ku_1\cos kt + u_2\sin kt}{u_2\cos kt +
ku_1\sin kt}}\right)-\displaystyle{\frac{1}{2k}}\\ [0.2cm]\\
\end {array}
$$
 {\bf Two Noetherian point symmetries}\newline
$$
\begin {array} {llcl}
L_{17}\quad &\Gamma_1 &\Rightarrow& -\displaystyle {\frac {1} {u_2\cos kt +ku_1\sin kt}}\\
[0.4cm] &k\Gamma_4+\Gamma_5 &\Rightarrow& 2\left(-\log\left
(u_2\cos kt +ku_1\sin kt\right) +1\right) \\ [0.2cm]\\ \hline\\
L_{18}\quad &\Gamma_3 &\Rightarrow& \displaystyle{\frac{1}{k}}\,
\log \left(
\displaystyle {\frac{ku_1\cos kt - u_2\sin kt }{ku_1\sin kt+u_2\cos kt} }\right)\\
[0.4cm] &\Gamma_6 &\Rightarrow&  -\log \left(-\half
(u_2^2-k^2u_1^2)\sin 2kt+ku_1u_2\cos 2 kt \right)\\ [0.2cm]\\
\hline\\ L_{34}\quad &\Gamma_3 &\Rightarrow&
-\displaystyle{\frac{1}{k}}\, \arctan
\left(\displaystyle {\frac{u_2}{ku_1} }\right)-t\\
[0.4cm] &\Gamma_4 &\Rightarrow&  \half \log \left(\displaystyle
{\frac{u_2^2+k^2u_1^2}{k^2}} \right)\\ [0.2cm]\\
\hline\\ L_{35}\quad &\Gamma_3 &\Rightarrow&
\displaystyle{\frac{1}{2k}}\, \log \left( \displaystyle
{\frac{-(u_2+ku_1)\cos kt + (u_2-ku_1)\sin kt }
{(u_2-ku_1)\cos kt+(u_2+ku_1)\sin kt} }\right)\\
[0.4cm] &\Gamma_5 &\Rightarrow&  \displaystyle{\half}\log
\left(-(u_2^2-k^2u_1^2)\cos 2kt -2ku_1u_2\sin 2kt \right)-1\\
[0.2cm]\\ \hline\end{array}$$
$$\begin{array}{llcl}
L_{47}\quad &\Gamma_7 &\Rightarrow& \displaystyle{\frac{1}{k}}\,
\arctan\left(\displaystyle {\frac{u_2}{ku_1} }\right) + t\\
[0.4cm] &\Gamma_8 &\Rightarrow&  \displaystyle{\frac{1}{2k}}\,
\log\left( \displaystyle{\frac{u_2^2+k^2u_1^2}{(ku_1\sin
kt+u_2\cos kt)^2}}
 \right)\\ [0.2cm]\\ \hline\\
L_{48}\quad &\Gamma_7 &\Rightarrow&
-\displaystyle{\frac{1}{2k}}\,\log\left(
\displaystyle{\frac{u_2^2+k^2u_1^2}
{(-ku_1\cos kt+u_2\sin kt)^2}} \right)+\displaystyle{\frac{1}{k}}\\
[0.4cm] &\Gamma_8 &\Rightarrow& \displaystyle{\frac{1}{k}}\,
\arctan\left(\displaystyle {\frac{u_2}{ku_1} }\right) + t\\ [0.2cm]\\ \hline\\
L_{57}\quad &\Gamma_7 &\Rightarrow&
\displaystyle{\frac{1}{2k}}\,\log \left( \displaystyle
{\frac{(u_2-ku_1)\cos kt+(u_2+ku_1)\sin kt}{(u_2+ku_1)\cos kt -
(u_2-ku_1)\sin kt }  }\right)\\
[0.4cm] &\Gamma_8 &\Rightarrow& \displaystyle{\frac{1}{2k}}\,\log
\left( \displaystyle {\frac{(ku_1\sin kt+u_2\cos
kt)^2}{\left[(u_2-ku_1)\cos kt+(u_2+ku_1)\sin
kt\right]\left[(u_2+ku_1)\cos kt - (u_2-ku_1)\sin kt \right]
}}\right)
\\ [0.2cm]\\ \hline\\
L_{58}\quad &\Gamma_7 &\Rightarrow&
\displaystyle{\frac{1}{2k}}\,\log\left(\displaystyle
{\frac{(u_2\sin kt-ku_1\cos kt)^2}{\left[(u_2-ku_1)\cos
kt+(u_2+ku_1)\sin kt\right]\left[(u_2+ku_1)\cos kt -
(u_2-ku_1)\sin kt \right] }}
 \right)\\
[0.4cm] &\Gamma_8 &\Rightarrow& \displaystyle{\frac{1}{2k}}\,\log
\left( \displaystyle {\frac{(u_2-ku_1)\cos kt+(u_2+ku_1)\sin
kt}{(u_2+ku_1)\cos kt -
(u_2-ku_1)\sin kt }  }\right)\\
[0.2cm]
\end {array}
$$

\section {Discussion}

In Sections 2 and 3 we provided the information necessary for the
discussion of properties of Lagrangians for (\ref {1.3}) given by
the method of Jacobi's Last Multiplier in conjunction with the Lie
point symmetries of the equation of motion (\ref {1.3}).  It is
apparent that Lie's method for the calculation of the Jacobi Last
Multiplier provides a direct route to the determination of many
Lagrangians for a system the equation of motion of which is richly
endowed with Lie point symmetries.  It takes no effort to infer
that even more Lagrangians could be obtained if one were to expand
the class of symmetries to include generalised symmetries.  The
disadvantage of this generalisation is that the calculation of
generalised symmetries for second-order equations is not as
algorithmic as is the calculation of point symmetries.  One must
make assumptions about the nature of the dependence of the
symmetry on the first derivative.  Another route which one can
take to obtain more Lagrangians is to make use of the fact that
the ratio of two multipliers is a first integral.  Once an
integral is obtained, further multipliers can be generated by
making use of the fact that an arbitrary function of an integral
is itself an integral.

An important consequence of the relationship, (\ref {1.2}),
between the last multiplier and a Lagrangian is that one obtains a
whole class of Lagrangians from a given multiplier.  Each member
of the class is an element of an equivalence class of Lagrangians
in the sense that each possesses the same number of Noether point
symmetries.  This is not necessarily the case for Lagrangians
obtained from different multipliers.  As we saw in \S 3,
Lagrangians leading to the same Euler-Lagrange equation can have
all possible numbers of Noether point symmetries.  Consequently
they are inequivalent Lagrangians.  In the classical case the
existence of inequivalent Lagrangians is a matter of some
curiosity.  Naturally it is more than merely curious if one
happens upon a Lagrangian which gives a lower number of Noether
point symmetries for then one could be misled on the subject of
the integrability of the underlying differential equation.

Although in this paper we have emphasised the approach of Lie for
the determination of the multipliers, we remind the reader that
the Jacobi Last Multiplier can be obtained in a variety of ways as
we indicated in the Introduction.  The method to be employed must
be gauged by the utility of the results which it produces.

\section*{Acknowledgements}

This work was undertaken while PGLL was enjoying the hospitality
of Professor MC Nucci and the facilities of the Dipartimento di
Matematica e Informatica, Universit\`a di Perugia.  The continued
support of the University of KwaZulu-Natal is gratefully
acknowledged.

\end {document}